\documentclass[prl,onecolumn,superscriptaddress,nofootinbib]{revtex4-1}
\usepackage{graphicx}
\usepackage{color}
\usepackage{amsmath}
\usepackage{amsfonts}
\usepackage{mathrsfs}

\begin{document}

\title[Phyllotaxis and packing]{Phyllotaxis: a non conventional crystalline solution to packing efficiency in situations with radial symmetry}

\author{Jean-Fran\c cois Sadoc}
\email{sadoc@lps.u-psud.fr}
\affiliation{Laboratoire de Physique des Solides (CNRS-UMR 8502), B{\^a}t. 510, Universit{\'e} Paris-sud, F 91405 Orsay cedex}

\author{ Nicolas Rivier }
\affiliation{ IPCMS, Universit\'e Louis Pasteur,  F-67084 Strasbourg, cedex }
\author{Jean Charvolin}
\affiliation{Laboratoire de Physique des Solides (CNRS-UMR 8502), B{\^a}t. 510, Universit{\'e} Paris-sud, F 91405 Orsay cedex}
\begin{abstract}
Phyllotaxis, the search for the most homogeneous and dense organizations of small disks inside a large circular domain, was first developed to analyze arrangements of leaves or florets in plants. Then it has become  an object of study not only in botany, but also in mathematics, computer simulations and physics. Although the mathematical solution is now well known, an algorithm setting out the centers of the small disks on a Fermat spiral, the very nature of this organization and its properties of symmetry remain to be examined.  The purpose of this paper is to describe a phyllotactic organization of points through its Voronoi cells and Delaunay triangulation and to refer to the concept of defects developed in condensed matter physics. The topological constraint of circular symmetry introduces an original inflation-deflation symmetry taking the place of the translational and rotational symmetries of classical crystallography.
\end{abstract}

\maketitle


\section{Introduction }

The  densest organization for a packing of small disks on an infinite plane is obtained when their centers are at the nodes of the triangular tiling of a hexagonal crystalline lattice, they all occupy the same area on this plane. If the disks are to be organized within a finite compact domain of the plane, this solution stays valid only if the boundaries of this domain are aligned along crystallographic directions. In the case of domain with a circular border, a constant area per center can only be obtained when the centers are regularly placed on the spiral drawn by the algorithm of phyllotaxis.

Phyllotaxis in order to tile   surfaces as a circular domain of the plane, a cylinder or a sphere leads to the building of distributions of points having the best homogeneity and isotropy possible\cite{bravais,jean1,jean2}. This solution appears in plant growth as being the only one compatible with a sequential growth within the frame of a self organizing process submitted to geometrical constraints \cite{turing,coxeter1,coxeter}. This has been noticed by D'Arcy Thompson\cite{darcy} who write: "...and not the least curious feature of the case is the limited, even the small number of possible arrangements which we observe and recognise."
 
Also for the optimization of homogeneity and isotropy, it appears  in the formation of B\'enard-Marangoni convection cells in cylindrical containers \cite{rivier,rivier2}, the organization of ferrofluid droplets falling down in silicone oil in presence of an inhomogeneous magnetic field with cylindrical symmetry \cite{douady} and that of air bubbles on a circular water surface \cite{yoshikawa}. Finally, spherical phyllotaxis was used to estimate the Earth coverage of satellite constellations \cite{gonzalez}.

A phyllotactic structure on a surface is a set of points which define position of physical objects as disks, convection cells, florets or others. We go into this matter describing a phyllotactic organization of points of the plane through its Voronoi cells and Delaunay triangulation and making call to the concept of defects developed in condensed matter physics. The topological constraint of circular symmetry introduces an original inflation-deflation symmetry taking the place of the translational and rotational symmetries of classical crystallography.

The phyllotactic structure is represented in Figs.~\ref{f2}, ~\ref{f3}. Outside a core, it consists of grains of hexagonal cells (in red) that are concentric circular rings, bounded and separated by circular grain boundaries  made of $n_2$ heptagonal cells (inner), $n_1$ hexagonal cells (middle) and $n_2$ pentagonal cells (outer). The two numbers $n_1$ and $n_2$ are successive Fibonacci numbers. These grain boundaries serve as natural boundaries for our optimal packing problem. Their properties which are presented  here therefore solved the optimal packing problem  topologically and metrically.

At the origin of our work are questions raised by the structure of collagen fibrils. They can be considered as dense packings with circular sections of more or less parallel rods, the so-called triple helices, whose lateral organization was the object of several investigations through X rays scattering studies. These observations could not agree with a purely hexagonal lattice and revealed the presence of an important degree of disorder. A lattice built from a distorted hexagonal lattice with a large unit cell, a quasicrystalline structure and paracrystalline fluctuations in a hexagonal lattice were then proposed. Those propositions are however difficult to conciliate with a twisted organization of the fibers, not compatible with the propagation of any long range lateral order, and the roundness of their normal section, as such orders should lead to  facetted sections. As some electron micrographs also suggest a spiral assembly, we recently examined a model in which the sections of the triple helices would be organized according to a phyllotactic pattern \cite{charvolinsadoc}.
Even if no definitive answer can be given, main features of the X ray diffraction patterns modeled using phyllotaxis are compatible with experimental results. We are actually investigating this problem.

\section{Phyllotaxis}

\subsection{The generative Fermat spiral}

\begin{figure}[tbp]
\includegraphics{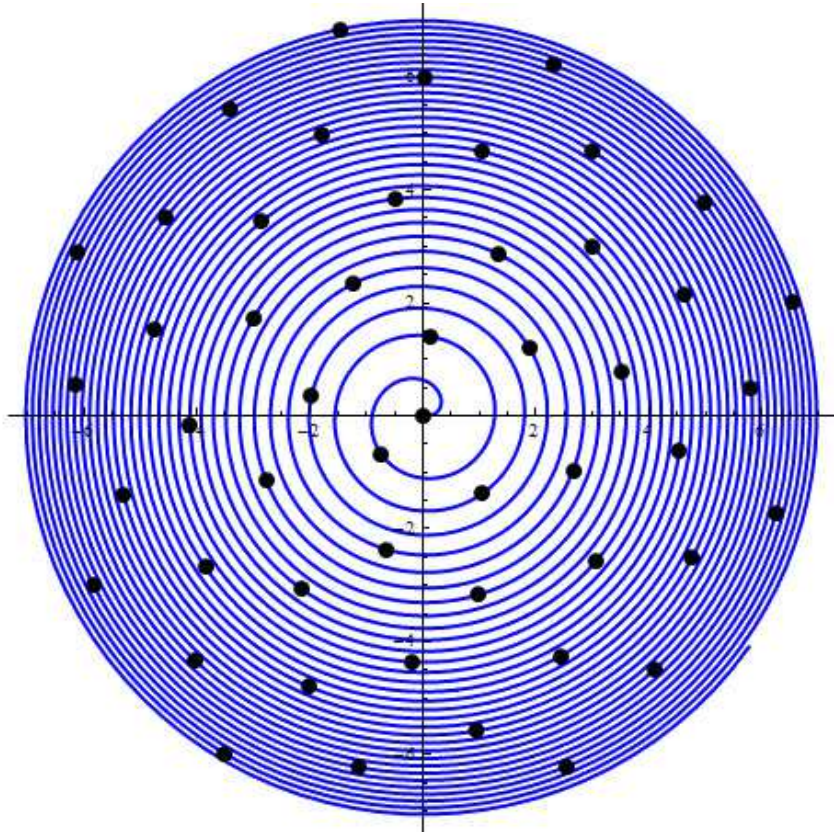}

\caption{A Fermat spiral with 50 points on it, successively placed at angle $\theta= 2\pi s \lambda$, with $s=0$ to $49$.}
\label{f1}
\end{figure}

Sites, the points at center of physical objects are lying on a  spiral  (figure~\ref{f1}) define by the equations relating cartesian coordinates $x,y$ with  polar coordinates  $\rho(s)=a \sqrt{s}$ and   $\theta(s)=2 \pi \lambda s$:
\begin{eqnarray}
\nonumber x(\rho,\theta)=a \sqrt{s} \cos(2 \pi \lambda s)\\
  y(\rho,\theta)=a \sqrt{s} \sin(2 \pi \lambda s)
\label{equa1}
\end{eqnarray}
where $a$ is a parameter defining the metric scale, $\lambda$ an important parameter which will be discuss below and $s$ the parametrization of the curve. Sites, indexed by integers $s$,  are placed on the curve, so that the azimuth between two successive sites  varies by $2 \pi \lambda $. The important result of the mathematical studies is that an homogeneous distribution of sites is only obtained with  $\lambda=1/\tau$ the inverse of the golden ratio $\tau=(1+\sqrt{5})/2$, or $\lambda=1/\tau^2$. The golden ratio is defined by
$1/\tau^2 -1/\tau -1=0$ so that this second  value gives the same result as the two corresponding angles have the same absolute value modulo $2\pi$ for integer $s$.

%
%
%
%

Two characteristic properties  of this generative spiral are important  in order to have the best covering. First,
the added area with any new site must be a constant, for any $\lambda$: this results of  $\rho(s) \propto  \sqrt{s}$,
but this choice is not sufficient to impose a local isotropy of the distribution of  sites. It is the second properties of the generative spiral which gives the best local isotropy of the area associated to the best uniform density: the choice of $\lambda=1/\tau$. With a rational $\lambda$, sites are aligned on straight radial spokes
and with irrational $\lambda$ different from $1/\tau$ (or of a noble number) they form a few long spiralling spokes, so in these two cases sites are gathered on lines \cite{ridley,rivier}.
Uniformity is indeed associated to the property of ``noble number'' to which $\tau$ and $\tau^2$ belong, to have only $1$ in the tail of their continued fraction expansion.  Then these numbers are approximated by successive truncations of their continuous fraction which converge smoothly \cite{adler98}. This point is discussed in \cite{rivier4}, which defines the best uniformity by the property of having shape and area of Voronoi cells as independent of $s$ as possible, a context-free inflatable structure; see also the so called ``shape invariance'' in \cite{rothen}.

We presents here  structures in which sites define local domains of the same area, but there are also examples where the area change with the radial position, like it is in a daisy where external florets are larger than those of the core. These organizations are obtained with other kind of spiral equations, for instance logarithmic spiral \cite{rothen}. Nevertheless conformal transformations but also shearing allow to explore these different phyllotaxis. This is discussed at the end of this text.

%
%
\begin{figure}[tbp]
\includegraphics{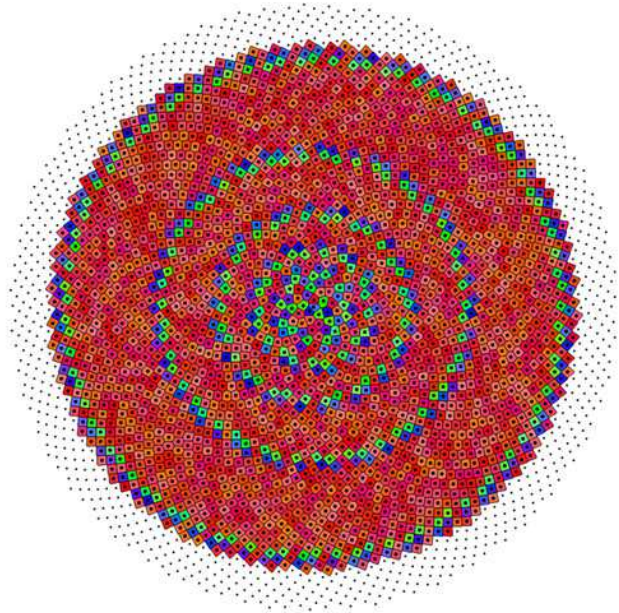}

\caption{Voronoi decomposition of a 2500 sites model. Voronoi cells define the number of neighbors: 6 for hexagonal cells in red, five for pentagonal cells in blue and seven for heptagonal cells in green. Small black points are for sites over 2500 added in order to avoid difficulties encountered to define Voronoi cells at the border of a finite set.}
\label{f2}
\end{figure}
%
%
\subsection{Voronoi tiling}

\begin{figure}[tbp]
\includegraphics{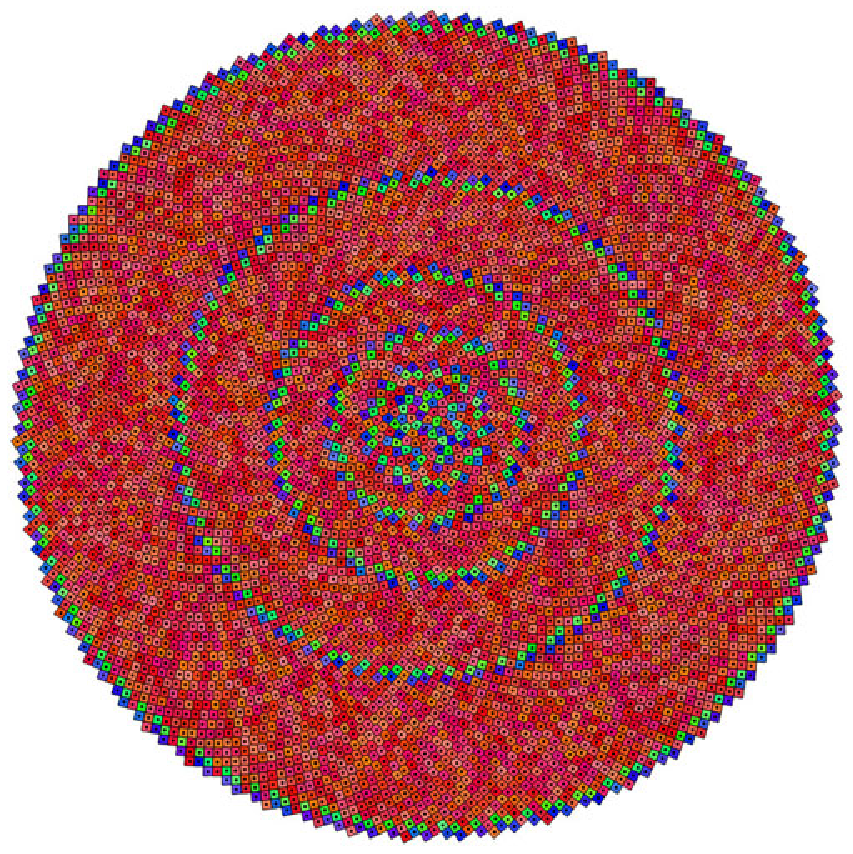}

\caption{Voronoi decomposition of a 6200 sites model.}
\label{f3}
\end{figure}

The geometry of a set of  sites can  be very well revealed using a Voronoi decomposition of the space (figures~\ref{f2} and~\ref{f3}). In 2D, each site is surrounded by a polygonal cell which is the locus of all the points of the plane closer to the considered site than to all other sites. In particular the Voronoi decomposition define strictly neighbors of a site whose cell share an edge with the cell of the considered neighboring site.
Building a Voronoi decomposition for a finite set of sites, we face the difficulty of defining neighbors close to the border of the set. We have solved this, by doing the decomposition for $N$ sites embedded in a slightly larger set.
With the generic hypothesis of a coordination number $c=3$ for Voronoi cell vertices, topological constraints are imposed by the Euler relation $F-E+V=\chi$, with the Euler-Poincar\'e characteristics $\chi=0$ for the infinite plane or for any closed surface without Gaussian curvature, like a torus. These constraints are such that  the average number of edges of Voronoi cells to be  strictly $6$.
But a phyllotactic pattern is a finite set with a circular boundary. For a compact finite part of the plane, the Euler relation  $V-E+F=\chi$ with the Euler-Poincar\'e characteristic  $\chi=1$ has to be considered.

In  phyllotaxis there are hexagonal cells, pentagonal cells and heptagonal cells only. Thus, a tiled plane surface enclosed inside a circle, in the limit  of an infinite radius must have the same number of   pentagonal  and heptagonal cells. It appears that the three kind of cells are organized by blocks.
Outside a core, it consists of grains of hexagonal cells  that are concentric circular rings, bounded and separated by circular grain boundaries $(f_{u-1}, f_{u-2}, f_{u-1})$ made of $f_{u-1}$ heptagonal cells (inner), $f_{u-2}$ hexagonal cells (middle) and $f_{u-1}$ pentagonal cells (outer). The $f_{u}$ are Fibonacci numbers defined by relation $f_u=f_{u-1}+f_{u-2}$ with $f_1=1$ and $f_2=1$ that is the sequence $1,1,2,3,5,8,13,21,34,55,89,144,...$. These grain boundaries serve as natural boundaries for our optimal packing problem. Outwards packing begins with the first complete grain boundary $(13, 8, 13)$ with 13 heptagons, 8 hexagons and 13 pentagons. The core disk is bounded by the 8 pentagons of the (first) incomplete grain boundary $(3, 5, 8)$. It has 3 heptagons instead of the full 8. The additional pentagons (nearly) fulfil the topological requirement that a tiled disk should have a topological charge of 6 (i.e. 6 additional pentagons, a sphere has a topological charge 12) \cite{rivier2,rivier3}.

 To resume, the hexagons are either in crystalline large grains (in a topological sense) or form  narrow rings
 between  heptagonal and  pentagonal rings.  The number of hexagons in the large rings are $...34,134,422,1221,3384,...$ as we shall see (table 1).

\begin{table}
\caption{\label{tab1}Table of cell types in the successive rings or annuli. Neighbor separations $\delta s$ are Fibonacci number, all the same for a given ring, except at the beginning. A label $u$ for large annuli corresponds to the Fibonacci number $f_u$ giving the medium separation ($f_u$ in the set of $(f_{u-1},f_u,f_{u+1})$ positive separations). Rings corresponding to defects are marked with $\|$. }
\begin{tabular*}{\textwidth}{@{}l*{15}{@{\extracolsep{0pt plus
12pt}}l}}
\toprule
$u$&Cell type&number of cells&$s$ from &to&neighbor separations  $\delta s$\\
\hline
&pentagon&2&0&1&1,2,3,4,5 or -1,2,3,5,8\\
&hexagon&1&2&2&-2,2,3,5,8,13\\
&heptagon&3&3&5&(-3,-2,2) or (-4,-3,-2) or (-5,-3,-2),3,5,8,13\\
&hexagon&1&6&6&-5,-3,3,5,8,13\\
&pentagon&2&7&8&-5,-3,5,8,13\\
&hexagon&1&9&9&-8,-5,-3,5,8,13\\
$\|$&hexagon&5&10&14&-8,-5,5,8,13,21\\
$\|$&heptagon&3&15&17&-13,-8,-5,5,8,13,21\\
$\|$&hexagon&5&18&22&-13,-8,-5,8,13,21\\
$\|$&pentagon&8&23&30&-13,-8,8,13,21\\
7&hexagon&2&31&32&-21,-13,-8,8,13,21\\
$\|$&heptagon&13&33&45&-21,-13,-8,8,13,21,34\\
$\|$&hexagon&8&46&53&-21,-13,-8,13,21,34\\
$\|$&pentagon&13&54&66&-21,-13,13,21,34\\
8&hexagon&34&67&100&-34,-21,-13,13,21,34\\
$\|$&heptagon&21&101&121&-34,-21,-13,13,21,34,55\\
$\|$&hexagon&13&122&134&-34,-21,-13,21,34,55\\
$\|$&pentagon&21&135&155&-34,-21,21,34,55\\
9&hexagon&134&156&289&-55,-34,-21,21,34,55\\
$\|$&heptagon&34&290&323&-55,-34,-21,21,34,55,89\\
$\|$&hexagon&21&324&344&-55,-34,-21,34,55,89\\
$\|$&pentagon&34&345&378&-55,-34,34,55,89\\
10&hexagon&422&379&800&-89,-55,-34,34,55,89\\
$\|$&heptagon&55&801&855&-89,-55,-34,34,55,89,144\\
$\|$&hexagon&34&856&889&-89,-55,-34,55,89,144\\
$\|$&pentagon&55&890&944&-89,-55,55,89,144\\
11&hexagon&1221&945&2165&-144,-89,-55,55,89,144\\
$\|$&heptagon&89&2166&2254&-144,-89,-55,55,89,144,233\\
$\|$&hexagon&55&2255&2309&-144,-89,-55,89,144,233\\
$\|$&pentagon&89&2310&2398&-144,-89,89,144,233\\
12&hexagon&3384&2399&5782&-233,-144,-89,89,144,233\\
$\|$&heptagon&144&5783&5926&-233,-144,-89,89,144,233,377\\
$\|$&hexagon&89&5927&6015&-233,-144,-89,144,233,377\\
$\|$&pentagon&144&6016&6159&-233,-144,144,233,377\\
13&hexagon&9167&6160&15326&-377,-233,-144,144,233,377\\
\hline
\end{tabular*}
\end{table}

%
%
%
%
%

%

\section{Rings and defects}

\subsection{Grain boundaries and defects}

\begin{figure}[tbp]
\includegraphics{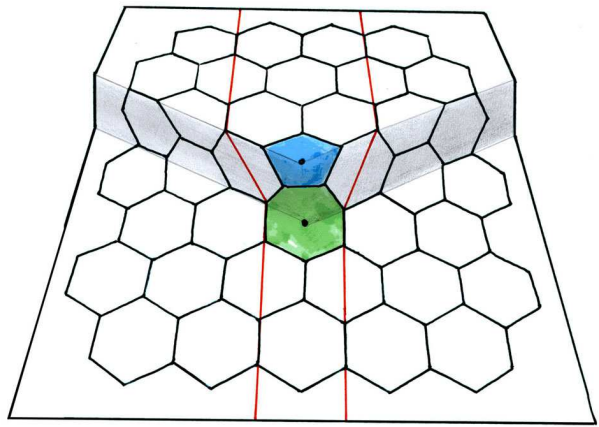}

\caption{A dislocation in an hexagonal crystal. The figure is drawn with regular polygons, so leading to the appearance of a stair. In 2D the structure is flatten by projection on a plane.}
\label{f4}
\end{figure}

In disordered material there are two kinds of disorder refereing to a perfect crystal. First is the  metric disorder such that relations of neighborhood  are conserved but with change for the length between neighbors, then topological disorder which imply defects like dislocations, disclinations breaking locally these relations. The decomposition of the phyllotactic pattern into rings gives good examples of these two kinds of disorder (figures~\ref{f2} and~\ref{f3}). The large rings of hexagonal cells have metric distortions but without defect. They can be seen as a ribbon cut into a perfect hexagonal crystal and then flattened onto a flat circular ring gluing its two ends, an operation which introduces   change of lengths (on a cylinder, it would be possible to wrap it without distortions). With reference to a perfect hexagonal 2D crystal, pentagons and heptagons are disclinations of opposite weights, but a dipole of opposite disclinations is a dislocation as shown on figure~\ref{f4}.  In the Voronoi decomposition pentagons and heptagons form dipoles which are dislocations in the structure. Several dislocations gathered along a line defined a grain boundary between disoriented grains. Hexagonal cells which are between two dipoles inside the dipoles ring, control the distance between dislocations, and then the disorientation between the two separated large hexagonal rings which are called crystalline grains even if they have metric disorder but without topological disorder. Dislocations are an important factor to kept the best homogeneity of the structure.
%
%
%
%

\subsection{Parastichies}

\begin{figure}[tbp]
\includegraphics{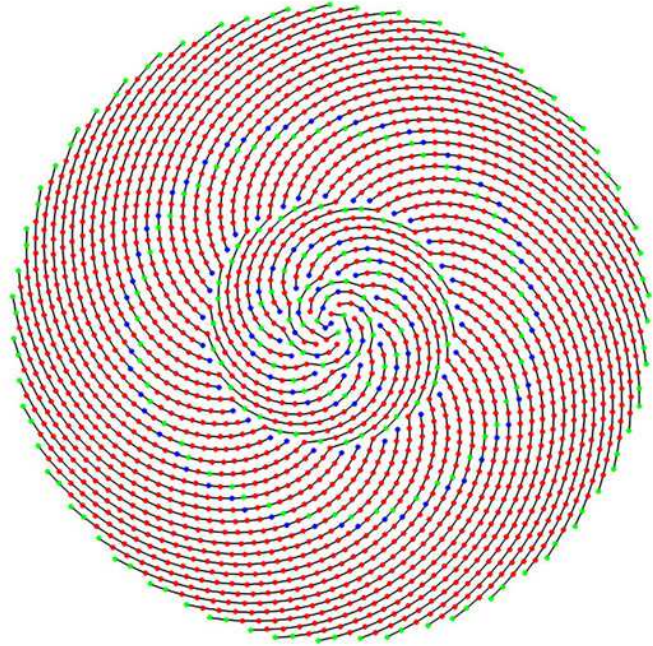}

\caption{Family of lines formed with parastichy segments, all rotating with the same orientation, corresponding to the smallest separation $\delta s$ given by a Fibonacci number $f_u$ with an even $u$ (for conservation of the rotation orientation).The color code is the same as on figure 2 and 3: red for 6 neighbors, blue for 5, green for 7. A new line originates when a new Fibonacci number, with these properties (the smallest with even $u$), appears in the list of separations with neighbors. Except for very small $s$, this happen on sites with pentagonal cells.These sites are gathered in one over two rings of pentagonal cells, such that in heptagon-pentagon dipole the new Fibonacci number is not in the list of separation for the site with the heptagonal cell, but the line is oriented as the dipole. All these lines play the role of reticular lines found in 2D crystals and clearly show the dislocations as an insertion of extra line.}
\label{f5}
\end{figure}

\begin{figure}[tbp]
\includegraphics{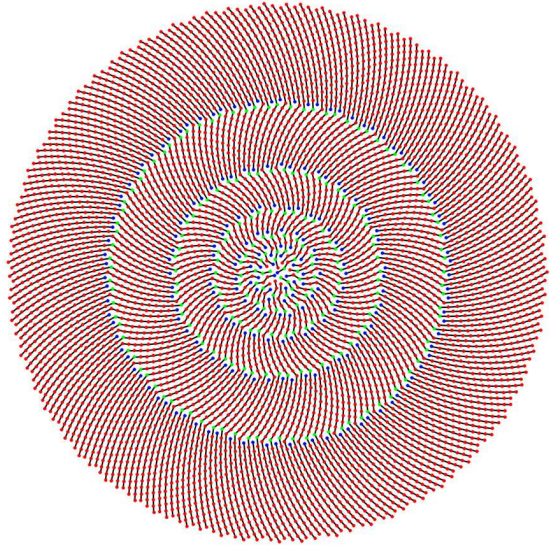}

\caption{Family of lines formed with parastichy segments,  corresponding to the largest separation $\delta s$ given by a Fibonacci number $f_u$.  Except for very small $s$, new lines originate on sites with pentagonal cells. The number of cell on the border is the number of these parastichies cut by it. }
\label{f6}
\end{figure}

Parastichy is a botanical term used to design  spiral lines observed on some plants such as composed flowers, cactuses and pine-cones. Looking at figures~\ref{f2} and~\ref{f3} such lines appear, even if they are not easy to follow all along. Some segments of parastichies appear on figures~\ref{f5} and~\ref{f6}. The purpose is now to describe them precisely. Points $s$ and $s+1$ placed on the generative spiral are separated by  long distances (at least for not too small $s$). For visible parastichies we are concerned by points which are neighbors, as defined by Voronoi cells, and search for the  separation $\delta s$ between two such points labeled  $s$ and $s+\delta s$.
Two points with polar coordinates $(\rho,\theta)$ are close if their   $\rho$ and their $\theta$ are close.
So $a \sqrt{s}$ must be close to $a \sqrt{s+\delta s }$, implying a small $\delta s/(2s)$. The two azimuthal separations must follow $2 \pi \lambda s- 2 \pi \lambda (s+\delta s)\simeq 0$ modulo $2\pi$. It is impossible to have in this equation a strict equality to zero because $\lambda$, the inverse of the golden ratio, is an irrational number  but  it can be approximated by a rational number $\lambda^\prime=f_{u-1}/f_u$, the ratio of two consecutive Fibonacci number. The equation for the azimuthal separation, with this approximation reduces to
$2 \pi \lambda^\prime s- 2 \pi \lambda^\prime (s+\delta s)=0$ which is solved exactly by $2\pi f_{u-1}=0 $  modulo $2\pi$ for $\delta s=f_{u}$. The approximation for $\lambda$ is better for large $u$, but as $\delta s/s $ must be small, first neighbors correspond to the best compromise between radial and azimuthal contribution.

Inside a ring of Voronoi cells of the same type, all neighbors are separated using the same set of Fibonacci numbers (except for $s<7$). For instance for the ring of hexagonal cells between $s=67$ and $s=100$ all six neighbors are at $s+\delta s$ with $\delta s$ in the set $\{-34,-21,-13,13,21,34\}$ of Fibonacci number $f_{9}=34, f_{8}=21, f_{7}=13$
(table~1).

Passing through a site of large hexagonal rings there are three parastichies defined by three consecutive Fibonacci numbers $f_{u-1},f_{u},f_{u+1}$ corresponding to three positive $\delta s$ separations.
It is convenient to label large hexagonal rings by $u$ such that $f_u$ correspond to the medium separation in this ring. So in the example with $67\leq s\leq100$ the label is $u=8$ corresponding to the Fibonacci number $f_8=21$ in the set $(13,21,34)$(see table 1).
%
%
%
%
%
%
%
\subsubsection{Equation of parastichies}

In a grain three visible parastichies are defined by three Fibonacci numbers $f_{u-1},f_{u},f_{u+1}$, but now we extend the notion of parastichies to a whole family of curves joining sites separated by any Fibonacci number.
So a parastichy is defined by a given Fibonacci number $f_u$.   We are showing that the parastichies equation has the form:
\begin{eqnarray}
\nonumber x(\rho,\theta)=a \sqrt{s} \cos(\theta_n+(-1)^{u+1} 2 \pi \lambda^{u} s/f_u)\\
  y(\rho,\theta)=a \sqrt{s} \sin(\theta_n+(-1)^{u+1} 2 \pi \lambda^{u} s/f_u)
\label{equa2}
\end{eqnarray}
The phase shift $\theta_n$ is adjusted in order to have the curve going through the site $s$, so $n$ is numbering all parastichies of a same family. Here we consider $s$ as a continuous variable, in order to define a curve, but we assure that the curve pass through sites for integer $s$ values.
Consider the two points $s$ and $s+f_u$ lying on a parastichies and on the generative spiral. On these curves their  azimuthal separation is $2\pi \lambda f_u$. This number which is largely greater than $2\pi$ can be reduced,
modulo $2 \pi$, to a small number between $-\pi$ and $\pi$ given by $2 \pi(f_u \lambda - f_{u-1})$ because $\lambda \simeq \lambda^\prime = f_{u-1}/f_{u}$.
The ratio $\lambda^u$ appears due to a property of power of the golden ratio: $\tau^u=f_u \tau + f_{u-1}$,
if $\lambda$ is the inverse of the golden ratio it follows \footnote{These two relations can be shown easily by induction considering that they are true then calculating  $\tau^{u+1}$ or  $\lambda^{u+1}$ using $\tau^2=1+\tau$ or $\lambda^2=1-\lambda$.} $\lambda^u=(-1)^{u+1}(f_u \lambda - f_{u-1})$.
 Introducing the expression for  $\lambda^u$, this azimuthal separation for $s$ varying from $s$ to $s+f_u$ is $(-1)^{u+1} 2 \pi \lambda^u$. Consequently the Fermat spiral $\{a \sqrt s \cos((-1)^{u+1} 2 \pi \lambda^u s/f_u +\theta_n ),a \sqrt s \sin((-1)^{u+1}2 \pi \lambda^u s/f_u +\theta_n )\}$ which rotates slowly compare to the generative spiral goes through the two points (with a good choice of the constant $\theta_n$) when $s$ vary continuously. Notice that  the scaling factor $a$ is the same for all spirals including the generative spiral.
A factor $(-1)^{u+1}$ changes the orientation of the spirals from even $u$ to odd $u$. This can also be put in relation with the fact that the approximation $\lambda \simeq f_{u-1}/f_{u}$ is an over estimation for odd $u$ and an under estimation for even $u$.
%
%
%
%
%

\subsubsection{A generalization of parastichies: reticular lines of the phyllotaxis}

In botany, parastichies are visible spirals of the structure and so are those joining close neighbor sites. In hexagonal rings there are three such lines characterized by three consecutive Fibonacci number, they are clearly visible when their characteristic Fibonacci number appears in  table 1.  Nevertheless for large $s$ even if they are less visible they always exist as continuous spirals joining sites not necessarily first neighbors, but second, third... neighbors. So to each possible Fibonacci numbers $f_u$  corresponds a family of $f_u$ identical spirals  joining sites with a separation  $\delta s=f_u$. There are $f_u$ in the family, because if one goes through the site $s$ and also through the site $s+f_u$, there is one through the site $s+1$ which is different, and one through $s+2$ and so on up to $s+f_u-1$.

 The first example is the family corresponding to $f_1=1$ with only one member,  the generative spiral joining site $s$ and $s+1$. The next example is given by $f_2=1$ also joining sites $s$ and $s+1$. These  are the two first families with a single spiral in each. In fact these two different spirals correspond to the two possible choices of $\lambda$ given by $1/\tau$ and $1/\tau^2$ giving the same structure with two spirals with reversed rotation orientations \footnote{It is very usual in literature to use $\lambda=1/\tau^2$ to describe phyllotaxis. This choice  corresponds to the smaller divergence angle. The other choice, done here is formally simpler and strongly supported by this property: spirals are defined by the rank $u$ of Fibonacci number; their speeds of rotation around the origin decrease with $u$ and we choose for generative spiral the first one given by $u=1$.}.
Exactly like for reticular lines in a 2D crystals  all sites are on a member of each family and lines of the same family are in some sense parallel lines: in crystals they are equivalent by translation, here they are equivalent by rotation. In 2D crystals, a reticular family is characterized by two Miller indices but the family of spiral is just characterized by  a Fibonacci number.
To resume, the number of spirals of the same family $f_u$ is the Fibonacci number $f_u$. As the golden ratio $\tau$ is approximated by the ratio of successive Fibonacci number converging toward it alternatively up and down, this result in an alternative rotation orientation for the spirals depending on the parity of $u$ in $f_u$.

\subsection{Euler relation}

As already said above, the Euler relation  $V-E+F=\chi$  imposes the average number of edges  per Voronoi cell to be 6 on the infinite plane for which  $\chi=0$. However, for a compact finite part of the plane, as is the circular domain of a phyllotactic pattern,  $\chi=1$ so that the number of cells with five or seven edges are not equal.
We introduce the number of faces $F=F_5+F_6+F_7$, $F_p$ being the number of faces with $p$ edges, and the number of edges $E=(5F_5+6F_6+7F_7)/2+(m_2+m_3)/2$ or in order to eliminate $V$, $E=3(V-m_2-m_3)/2+(2m_2+3m_3)/2$. The number $m_2$ and $m_3$ are the number of vertices of the cells on the border with coordinations $c_r=2$ or $3$. This gives
$(m_2-m_3)+(F_5-F_7)=6$. Two particular solutions of this equation are interesting. First, following what happen on the infinite plane, suppose  that $F_5=F_7$ so there is $m_2=6+m_3$, an excess of six sites with coordination $2$ on the border. A simple  example is obtained with a finite piece of an hexagonal tiling cut along edges of hexagonal cells ( so with $F_5=F_7=0$). Then this structure has an irregular boundary depending on the choice of hexagons kept inside it respecting $m_2=6+m_3$ for the $m_2$ sites of the boundary belonging to only one hexagon, and the $m_3$ sites belonging to two hexagons. The other particular solution, corresponds to the phyllotaxis: the boundary is close to  a circle with  $m_2=m_3$, an alternative equipartition of coordination $2$ and $3$ on the border. Notice that in a perfect hexagonal tiling  cut along edges in order to divide it in two pieces with sites on the cut alternatively on one or two hexagons ($c_r=2$ or $3$), the cutting line oscillates around a straight line, so we can consider such border as some kind of ``topological  geodesic'' \cite{rivier3}. In this case, solving the equation,  gives $F_5-F_7=6$ corresponding to an excess of six pentagons in the core of the structure. Evidently a circular border with these properties ($m_2=m_3$ forcing $F_5-F_7=6$) must be inside large hexagonal domain and not inside rings of defect: otherwise there would have an excess  of heptagon. A consequence of $m_2=m_3$ on the border, is also that the number of cells on the border (having outside edges) does not varies continuously with the total number of site in the phyllotactic pattern: this number remains constant in large hexagonal rings, but otherwise  changes rapidly when the boundary enter inside defect rings. This number  is equal to the number of parastichies corresponding to the highest separation between neighbors, so if the border is inside non-defective domain it   is equal to   Fibonacci numbers.
%
%
%
\begin{figure}[tpb]
\caption{Variation of the area of Voronoi cells for points between $s=34$ and $2166$. Colors correspond to the type of Voronoi cells. The area are close to the mean value $\pi$ with $a=1$, strong variations appear in rings of defects. Nevertheless these strong variations decrease as the area tends toward $\pi$ for large $s$.}
\includegraphics{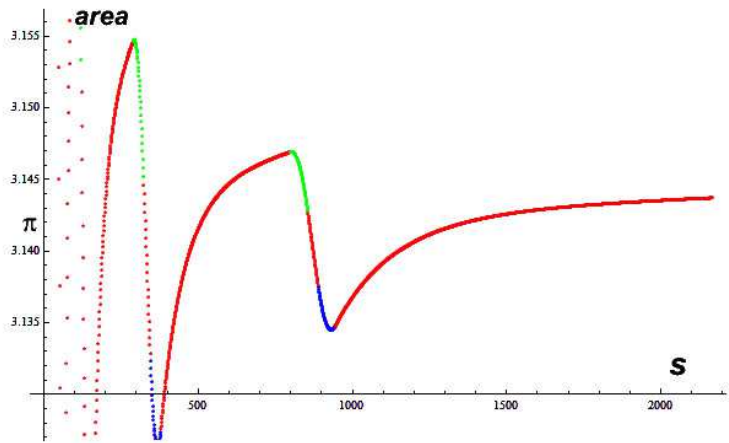}
\label{f7}
\end{figure}
%
%

\subsection{Rings and defects: a relationship with crystals}

Exactly like for reticular lines in a 2D crystals  all sites are on a member of each family and lines of the same family are in some sense parallel lines: in crystals they are equivalent by translation, here they are equivalent by rotation. In 2D crystals, a reticular family is characterized by two Miller indices but the family of spiral is just characterized by  a Fibonacci number.
To resume, the number of spirals of the same family $f_u$ is the Fibonacci number $f_u$. As the golden ratio $\tau$ is approximated by the ratio of successive Fibonacci number converging toward it alternatively up and down, this result in an alternative rotation orientation for the spirals depending on the parity of $u$ in $f_u$.

There are relationships between crystals with defects and phyllotactic pattern, but in a next section relations to quasicrystals are given.

\section{Metric properties}

\subsection{Packing density and fluctuations}

As the circle of radius $\rho$ and area  $\pi \rho^2=\pi a^2 s$ contains $s$ points, the area added per point has  the  value
 $\pi a^2$.The area of Voronoi cells   oscillates close to this value for small $s$ and then converges towards it (fig. \ref{f7}).

%
%
%
\begin{figure}[tbp]
\includegraphics{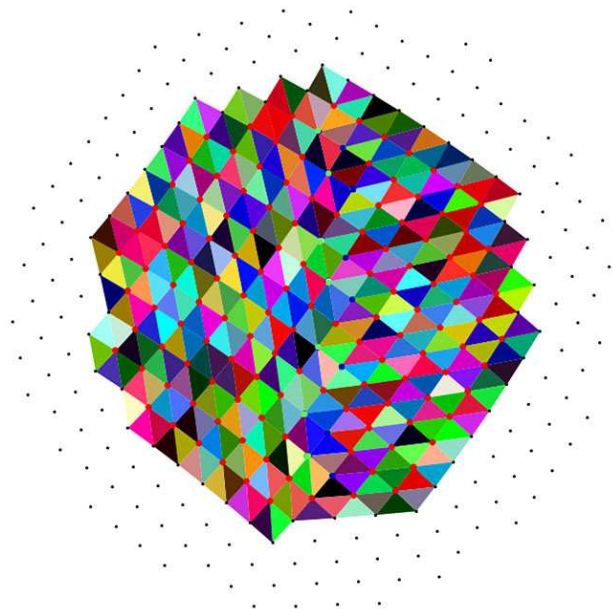}

\caption{Delaunay triangulation in a small region around the point $s=15326$ which is the last point in a wide six neighbors ring. The center of the phyllotactic disk is far away on the left. Red points are common to 6 triangles, blue points to 7 triangles and green points to 5 triangles. Often it seems that two triangles form a square, but in fact in this sheared square the diagonal which divide it into triangles is slightly shorter than the other one. There are two domains differentiated by the diagonals orientation. In these domains all points are common to 6 triangles. Between these two domains the defect ring appears as a place where some diagonal orientations are flipped, leading to 7,5 or eventually 6 triangles around a point. Going away from the defect domain, triangles become more and more regular.}
\label{f8}
\end{figure}
%

\subsection{Influence of site distances on shapes of Voronoi cells}

In the Voronoi decomposition of a phyllotactic pattern there are only hexagons, pentagons and heptagons, nevertheless, looking rapidly at figure 3 some squares seem to appear, mainly in rings related to dislocations. In fact these are not squares, but   Voronoi cells which   could be described as a square with one corner slightly truncated (for pentagon), or with two truncated corners (for hexagon), or three truncated corners (for heptagon).
This is to relate to the fact that the two parastichies corresponding to the two first Fibonacci separations are close to be orthogonal in rings of defects. Oppositely in large rings of hexagonal cells, the core of the ring contains less irregular hexagonal  cells.
 %
%
%
\begin{figure}[tpb]
\caption{First neighbor distances between sites $s$ and $s+\delta s$. Blue, for the interval $\delta s$ equal to the smaller positive Fibonacci number in the list of table 1. Green, for the next interval, red for the third positive interval and purple for the last one (occurring only if the  Voronoi cell is an heptagon). So  green, blue and red correspond to distance along the three visible parastichies. Each continuous curve corresponds to a given Fibonacci number that appears in different annuli.
For instance, $f_{11}=89$ appears between $s=290$ and $5926$ leading to a continuous curve which is successively purple, red, green and blue. Each lower crossing of two curves corresponds to a grain boundary, upper crossing is in the middle of a hexagonal annulus.
 The scaling parameter in equation (1) is $a=1$, so the distances are approximatively in the range $[1.676, 2.506]$ with a mean value $1.903$. }
\includegraphics{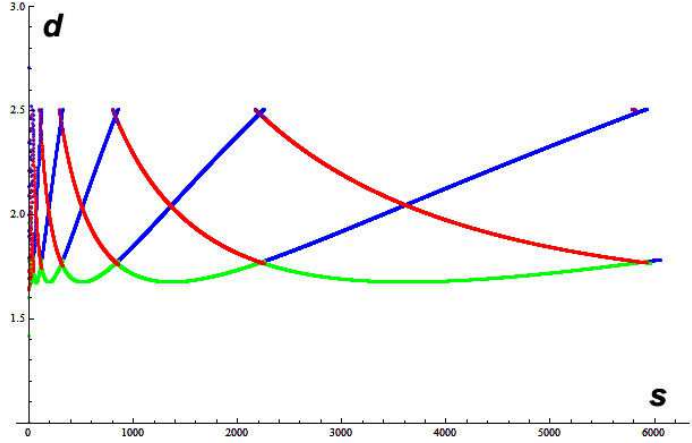}

\label{f9}
\end{figure}

Figure \ref{f8} shows how Delaunay triangles are organized near the rings of defects \footnote{Delaunay triangulation is such that sites are associated to cover the surface by triangles. It is related to Voronoi decomposition, because vertices of Voronoi cells are centers of the triangles, Delaunay and Voronoi decompositions are dual.}. In these regions corresponding Voronoi cells are close to be square, but with corner slightly truncated. Effectively Voronoi cells vertices are centers of circumcised circles of Delaunay triangles, so if two such triangles look like a square divided by a diagonal the two centers are close to the diagonal leading to a small edge of the Voronoi cell.   If two corners are truncated the cell is an hexagon. In large hexagonal rings, the Voronoi cells are  rhombus with two opposite truncated corners, so the Voronoi cells are close to regular hexagons. In the narrow hexagonal rings, inside defect rings, the cells look like squares with two close truncated corners, so are very distorted hexagons.

Even if Delaunay and Voronoi decomposition give exactly the same information some properties are more easily observed with one or the other method. For instance angles between parastichies are very clear on the figure \ref{f8} because edges of triangles are along parastichies. It is clear in this region of defects that two families of parastichies are close to be orthogonal with a third family making a $\pi/4$ angle with the others.

\subsection{Distances between first neighbors}

The behavior of distances  between first neighbor sites is shown on  figure \ref{f9}.
In domains where Voronoi cells are slightly deformed squares, their area is approximatively $\pi a^2$ so that the distance between points, along a square edge, is  $\sqrt{\pi}\simeq1.772$ (with $a=1$) and along a diagonal, is $\sqrt{2\pi}\simeq2.506$.
 When Voronoi cells are more clearly hexagons two distances are close to $\sqrt{\frac{3\pi}{\sqrt{5}}}\simeq2.053$.   The behavior of the  distance between the two points defined by $s$ and $s+f_u$  as a function of $s$ is obtained using relation given by Yeatts~\cite{yeatts}:
 \begin{equation}
d_u(s)=f_u (\frac{1}{4s}+\frac{ s (-2\pi f_{u-1}+2\pi \tau^{-1}f_u)^2}{f_u^2})^{1/2},
\label{equa3}
\end{equation}
this relation perfectly fits the numerical values given on the figure~\ref{f4}. It can be checked that minimal values are
\begin{equation}
\sqrt{2 \pi } f_u \sqrt{|\frac{1}{\tau }-\frac{f_{u-1}}{f_u}|}
\label{equa3b}
\end{equation}
  all close to and converging toward $\sqrt{\frac{2\pi}{\sqrt{5} }}\simeq1.67$.

 All distances are in the range $[\sqrt{\frac{2\pi}{\sqrt{5} }},\sqrt{2\pi}]$, whatever will be $s$ and $u$.  This property, specific to the choice of $\lambda=1/\tau$ is very important to ensure the best packing homogeneity
 \footnote{These limits remain the same in phyllotactic tilings on the sphere and on the hyperbolic plane, providing the same $\pi$ area for each site. This will be given in a further article on phyllotaxis in non-Euclidean geometry.}.

 \section{grain boundaries}

Table 1 shows lower and upper bounds of domains containing the same kind of Voronoi cells. These values are obtained from numerical observations of a phyllotaxis. Nevertheless it would be interesting to have an analytic relation giving these bounds. Clearly having the lower bound for heptagonal or pentagonal rings the upper bound is obtained adding a Fibonacci number minus one. This is also true for narrow hexagonal domain, in the same region, using the previous Fibonacci number. But there is no clear relation for wide hexagonal rings. We introduce a semi-empiric relation which has been tested up to $s=280783$.

\begin{figure}[tbp]
\caption{Strip cut in a square lattice (a). This strip is a good approximation of the ring of defects between points $s=101$ and $s=155$ containing $55$ points. Color of the points correspond to the type of their Voronoi cells. Point $s=101$ is the first green point down, then points are numbered, increasing by $21$ going up or decreasing by $34$ going right. This strip can be divided into three strips of heptagons, hexagons and pentagons. (b and c) Continuation of the grain toward the other boundary  on the same square grid $\{(x,y):x-1\leftrightarrow s+34, y-1\leftrightarrow s+21\}$ which can be considered as folded on a cylinder with axis perpendicular to the strip. This grain (in grey) is limited by an inner border (b) and an outer border (c). Blue points (pentagons) in (b) are those of (a), orange points belong to the large grain. The grain is bordered in (b) by a Fibonacci chain defined by the lower part of the grey domain. There are 21 stairs in it containing 21 blue points and  34 orange points (hexagons of the large grain). The other border of the grain (in (c)) is geometrically identical, but it contains 34 green points (heptagons of the next ring of defects) and 21 red points (hexagons of this ring of defects). This next ring of defects from $s=290$ to $s=378$ contains 89 points. The beginning of this ring is expanded in (d), with two Fibonacci chains, one which close the grain from $s=156$ to $s=289$ and the other which begin the next grain,  from $s=379$ to $s=800$. The two Fibonacci chains could be coded with long (purple) and short (blue) segments forming a ``stair case''. Stairs of the first chain  appear with $\pi/2$ angles, but for the other angles are $\pi/4$.
In fact the strip should be deflated (stretched along by a rational approximant of $\tau$ and compressed across by an approximant of $1/\tau$, that is a Poisson shear). Introducing progressively the Poisson shear, instead of a part of a cylinder the substrate of the grain and its two boundaries is a part of a surface with a negative constant Gaussian curvature around the same axis. This Poisson shear change the rhombic cell darken in (d) into a square. Refer also to figures 11 and 12.}
\includegraphics{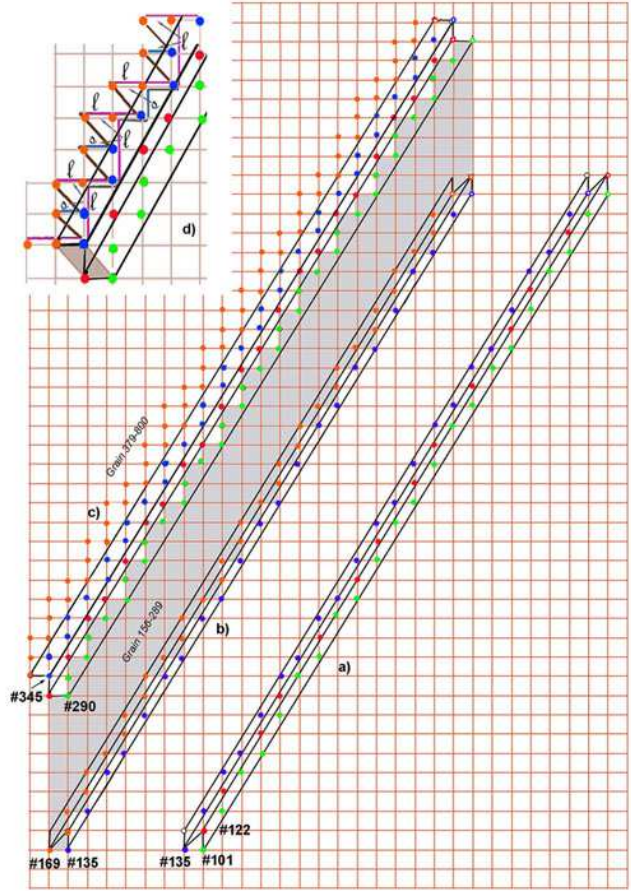}
\label{f10}
\end{figure}

As in a ring of defects Voronoi cells are close to be squares, we suppose that  sites are with a good approximation on a square lattice with parameter $d=\sqrt{\pi}$. We set $a=1$ in the following.
Figure~\ref{f10}-a shows a representation of the defect ring at $101\leq s\leq 155$ in which Voronoi cells are 21 heptagons,13 hexagons and 21 pentagons. It appears straight in this representation, but in the phyllotaxis it is refolded into a ring.  The defect ring appears as a strip of length given by the vector $(21,34)$ in the square lattice and a width given by $(1,-1)$, so  that the number of points in the strip is $55$, given by the determinant constructed with  the two vectors. This strip is separated into three thinner strips of same length, but with widths defined by vector $(0,-1)$, $(1,1)$ and $(0,-1)$ respectively. The first one contains the 21 points with heptagonal  cells, the second contains 13 points with hexagonal cells and the third contains 21 points with pentagonal cells \footnote{Notice that this strip is very similar to the construction of approximants of quasicrystal using the cut and projection method. It is one of the reasons which relate phyllotaxis and quasicrystal.}. In the phyllotactic pattern, the 55-points strip is folded into a flat ring that bounds a disk, by identification of the two small sides at the price of metric distortions.
  %
%
%
%
%
 We estimate the bounds  of the strip using two properties: the length of the strip and the number of points in the disk that is enclosed with radius $\sqrt s$. The length is the module of the vector $(21,34)$ which is $L=\sqrt{21^2+34^2}\sqrt{\pi}$ using $d=\sqrt{\pi}$ as length unit. Folding  the strip into a ring, this length is the perimeter of a circle. We can estimate the number of points of the phyllotactic pattern in the disk enclosed by this circle using the definition of the generative spiral. This number is $s_b=\frac{21^2+34^2}{4 \pi}\simeq 127.085$.

 Now we compare this to the known value for upper and lower bounds in this rings of defects. When the strip is folded into a ring,  the internal side of the ring is compressed whereas the external side is expended, so  $s_b$ is collimated by  the narrow ring of hexagons. The evaluated value $s_b$ falls between these  bounds  $(122,134)$.

 From this estimation we  deduce a formula which gives bounds of all hexagonal rings appearing inside defects regions:
 \begin{eqnarray}
 \nonumber({\rm round}[\frac{3-f_{u-1}}{2}+\frac{f_u^2+f_{u+1}^2}{4 \pi }],{\rm round}[\frac{f_{u-1}+1}{2}+\frac{f_u^2+f_{u+1}^2}{4 \pi }])
\end{eqnarray}
It is possible to write this relation in a more compact way using the identity $f_u^2+f_{u+1}^2=f_{2u+1}$ , which can be shown by inductive reasoning on $v$ using $f_{u+v} = f_{v+1}f_{u}+f_{v}f_{u-1}$. Then we have:
\begin{eqnarray}
 ({\rm round}[\frac{3-f_{u-1}}{2}+\frac{f_{2u+1}}{4 \pi }],{\rm round}[\frac{f_{u-1}+1}{2}+\frac{f_{2u+1}}{4 \pi }])
   \label{equa4}
\end{eqnarray}
Here $f_u$ is the Fibonacci number of order $u$ and the function round take the integer part of an irrational number.  The example of the figure \ref{f10} corresponds to $f_8=21$. In this relation the two bounds are such that the number of point in the ring is $f_{u-1}$ as wanted. This formula has been checked to give exact value (except a shift of one for $u=11$)
up to the bounds $ (280174, 280783)$ obtained with $u=16$. Using the fact that rings involved in defect domains have number of points given by Fibonacci number it is easy to get all other bounds as given in table 1.

There is an other way to look at this problem. It is to search $s$ for points where two parastichies are very close to be orthogonal. This relation which is deduced from results given   by Yeatts \cite{yeatts}  is:
$ s=\sqrt{{\rm abs}[\frac{f_{u} f_{u+1}}{4 \gamma_u \gamma_{u+1} }}]$ where $\gamma_u$, related to approximants of $\tau$, is $\gamma_u=2\pi (\lambda f_u-f_{u-1})$.
This relation give real values for $s$ very close to $ \frac{f_{2u+1}}{4 \pi }$ but clearly different (for instance $s^{\prime}_b\simeq127.067$ compared to $s_b\simeq127.085$).

It will be a challenge to found an exact relation giving integer values for the bounds without ambiguity on the use of function  ``round'',``floor'' or ``ceiling'' extracting integers from non integer numbers.

The ratio between  radius of two successive  rings of defects, are proportional to  $\sqrt{f_{2u+1}/f_{2u-1}}$ very close to the golden ratio.

\section{Conformal transformations and shearing in phyllotaxis }

In this section, we show how the successive grains are sheared relatively to a perfect crystalline structure, so that each parastichy is a stack of deformed hexagons, starting from square at one boundary, through more symmetrical hexagon at mid-grain, to square at the next boundary. See figure~\ref{f9}, where the lowest crossings correspond to square cells at the grain boundary with $d = \sqrt{\pi}$ (the third distance is the diagonal $\sqrt{2\pi} \simeq 2.507$, the maximal distance). The upper crossings correspond to  hexagons with $2\pi/3$ angles (but not perfectly regular) at mid-grain, with a minimal distance given by (\ref{equa3b}).

Phyllotaxis (1) is an example of {\it spiral lattice} \cite{rothen2}, the image $D$ in the complex plane $ {\bf w} = \rho \exp {i \theta}$ of a regular lattice $P$ in a domain of the complex plane ${\bf z}$ with cartesian coordinates. In elasticity, $D$ is the actual, deformed state of the material, and $P$ is called the natural state. The mapping is through the function ${\bf w} ({\bf z}) $.

One particular mapping ${\bf w} ({\bf z}) = {\bf w_0} exp (\bf{b^*.z}) $ is conformal, ie. it is analytic, invertible, singularity-free (excluding  a small domain around the origin of ${\bf w}$), and thus without any defect (dislocation, disclination or grain boundary). It conserves the angles and is also unsheared (a consequence of the Cauchy-Riemann relations imposing analyticity)  \cite{rothen2}.

We   consider a square lattice and we define its nodes positions by complex numbers in its plane. As an example we consider a strip in this lattice, like that shown on the figure~\ref{f10} with square vertices selected in. We have to map it on a ring of dipole. On this example the strip is define by a long multiple cell of the square lattice characterized by the two vectors ${\bf b_2} =(21,34)$ and ${\bf b_0}=(1,-1)$. The conformal mapping transform the imaginary axis into a circle, so we have to choose this axis along the ${\bf b_2}$ vector. As complex numbers are defined using orthogonal real and imaginary axes, we introduce the real axis along the vector ${\bf b_1}=(34,-21)$. To express the argument of the mapping function we introduce reciprocal vectors of ${\bf b_1}$ and ${\bf b_2}$ called ${\bf b_{1}^*}$ and ${\bf b_{2}^*}$ and defined by ${\bf b_i}.{\bf b_{j}^*}=\delta_{ij}$. Then the mapping function is
${\bf w} ({\bf z}) = {\bf w_0} exp[2\pi( {\bf b_{1}^*} +i {\bf b_{2}^*})(x{\bf b_1} +y {\bf b_2})] $ where $x$ and $y$ are coordinates of a point ${\bf z}$ expressed in the multiple cell. The mapping function reduces simply to ${\bf w} ({\bf z}) = {\bf w_0} exp[2\pi(x +i y )] $ with the scaling parameter ${\bf w_0}$

The square lattice on $P$ includes  all points with $s \geq 101$. Note the particular families of reticular lines, (i) (i) $(1,0)$, $(0,-1)$ and $(1,-1)$, that are mapped into parastichies, (ii) $(21,34)$, parallel to the grain boundary, that are mapped into concentric circles, and (iii) $(34,-21)$ that are mapped into radial spokes. The conformal mapping  yields a spiral lattice \cite{rothen2}, with all reticular lines mapped into equiangular (or logarithmic) spirals, save $(21,34)$ (concentric circles) and $(34,-21)$ (radial spokes). All angles are conserved in the mapping, notably the two parastichies, images of $(1,0)$ and $(0,-1)$ remain at right angles and keep the same angle with the circles parallel to the grain boundary and with the perpendicular radial spokes.
Thus, figure~\ref{f10} is a conformal (literally) representation of the grain boundary, with square-shaped Voronoi cells. However, the density of points $|dw/dz|^{-2} = |b w|^{-2}$ is not uniform and the conformal mapping must be sheared, from the grain boundary outwards, to achieve the required uniformity. Accordingly, the hitherto defect-free phyllotaxis will exhibit other grain boundaries further out. Remarks: a) the mapping is enacted outwards, as occurs naturally in the growth of phyllotactic structures, younger florets pushing out the older ones and squeezing in-between at grain boundaries;
b) we can say nothing for small $s$. In particular, on the central core of the phyllotaxis $s < 32$ that is probably the smoothest way of accommodating the core of a weak, non-singular topological defect that is here extra-matter \cite{rivier3}.

Figure~\ref{f10}--a is a representation of the lattice $P$. It consists of adjacent, non-overlapping parallel strips of equal thickness generated by the translation ${\bf b_1}$. Each strip in bounded by two reticular lines parallel to $(21,34)$ and contains 55 points, the diagonal of the unit square goes as the thickness of the strip. The two points are separated by $\delta s = 55$. The first strip is the grain boundary itself. Successive strips are mapped conformally by
 ${\bf w} ({\bf z}) = {\bf w_0} exp[2\pi(x +i y )] $ into concentric rings containing 55 points each, but with thickness increasing exponentially $\rho (s+55) = \rho(s)\exp(55)$, where $\rho = |w|$. In order to have uniform density of sites in the presented phyllotaxis with $\rho(s) = a \sqrt{s}$, the successive strips in $P$ must be squeezed  across and extended through a Poisson  shear strain (figure~\ref{f10}--d).

\section{Relation with quasi periodic structures}

The way dislocation dipoles are organized along circles is strongly related to 1D quasicrystals. A Fibonacci 1D quasicrystal can be obtained using an inflation-deflation rule iteratively applied to a sequence of long and short segments. This  inflation-deflation rule is $L \rightarrow L + S$ and $S\rightarrow L$. Starting simply from a short segment this specific rule gives a quasicrystal after an infinite number of iterations, or with a given number of iterations, a finite structure with a number of short and long segments given by two successive  Fibonacci numbers.  Consider now dipoles along circles in the phyllotaxis. There are isolated dipoles (an heptagon and a pentagon in contact) and pair of close dipoles. Consider now the rule: change isolated dipole into a pair of dipoles and change pair of dipoles into a pair  and an isolated dipole. It is the rule we have going to one ring of dipoles to the next ring. So there is an inflation-deflation symmetry associated with radial change relating defects in this structure. This can be checked counting the number of pair of dipoles or of isolated dipoles which are successive Fibonacci numbers, on circles of defects. The figure \ref{f10} which is similar to figures used in presentation of one dimension quasicrystal defined by cut and projection is related to this property.

%

\subsection{Successive grain boundaries}


The first complete grain boundary $(13,8,13)$ marks the beginning of a regular, inflatable phyllotaxis, where the circular grain boundaries are words, sequences of letters $L$ and $S$, where $L$ labels a dislocation (oriented dipole 7/5) and $S$ an isolated hexagon in the grain boundary. The sequences are obtained by inflation rules $S \rightarrow L$, $L \rightarrow LS$. Consider three successive grain boundaries, with sequences $w_{i-2}$, $w_{i-1}$ and $w_{i}$. Inflation implies that the sequence $w_{i}$ is given by concatenation of the previous two, $w_{i} = w_{i-1} \bullet w_{i-2}$, thus $w_{i} = LSLLSLSL $ is a concatenation of $w_{i-1} = LSLLS $ with $w_{i-2} = LSL $. The word $w_{i}$ has $|w_{i}| = f_{i}$ letters, $f_{i-1}$ $d$ and $f_{i-2}$ $s$. It starts with $LS$ and ends with $LS$ for $i >3 $ with even $f_i$, with $SL$ for $i >3$ with odd $f_i$.
A circular word is also a palindrome. Let $w^T$ be the sequence $w$ read backwards (transposed in matrix notation). A palindrome $p = p^T$ is sequence of letters that reads the same either way, e.g $LSLSL$, $hannah$ or $eve$. The word $w_{i}$ can also be written as the concatenation of two palindromes $w_{i} = p_A \bullet p_B$, with $w_{i-1} = p_A SL , p_B = SL w_{i-2}$ for $i$ even, and $w_{i-1} = p_A LS , p_B = LS w_{i-2}$ for $i$ odd. Thus $|p_A| = f_{i-1} - 2 , |p_B| = f_{i-2} + 2$. For inflation to be consistent from one grain boundary to the next, one requires $|p_A| > |p_B|$, so that the shortest complete grain boundary is $(13, 8, 13)$, a palindrome of 21 letters $L$ and $S$, written on a circle and readable in either sense, since $w = p_A \bullet p_B = p_B \bullet p_A $ (by circular invariance) equals $w^T =  p_{B}^T  \bullet p_{A} ^T $.

Thus, $p_A^{i} = p_A^{i-1} LS p_A^{i-2} = p_A^{i-2} [SL] p_A^{i-3} [LS] p_A^{i-2}$ for $i$ even, and $p_A^{i} = p_A^{i-1} SL p_A^{i-2} = p_A^{i-2} [LS] p_A^{i-3} [SL] p_A^{i-2}$ for $i$ odd, with $p_A^{3} = 1$
%

\subsection{Boundary of large hexagonal rings}

A large hexagonal ring which is lying on the  plane is nevertheless topologically equivalent to a perfect hexagonal tiling wrapped around a finite cylinder limited by two parallel circles. With such distortion of the ring, hexagonal cells are perfect regular hexagons and so this geometry can be seen as a multiple crystallographic cell of the hexagonal lattice with two identified opposite sides. The  vectors which are the basis of the multiple cell entirely characterize the topology of its related large hexagonal ring. In order to precisely define the boundaries of a ring we will consider in it all pentagonal sites which are just before and all heptagonal sites just after  associated with the following narrow hexagonal ring. With this definition of large ring boundaries the set of all large rings entirely cover continuously  the whole structure. They are perfect topological hexagonal structures each defined by a crystallographic multiple cell which multiplicity is given by the number of points in it. The departure from the hexagonal coordination only results from the junction between successive rings.
 %
%
%
\begin{figure}[tbp]
\includegraphics{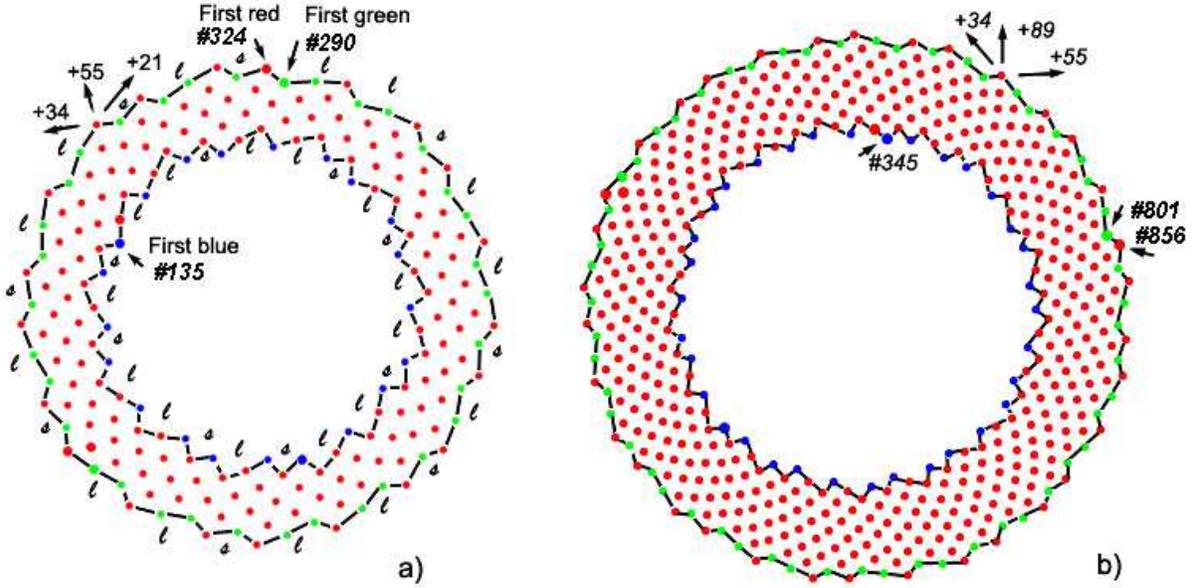}

\caption{Two example of hexagonal rings: a) with 134 hexagonal sites (red), completed with 21 pentagonal sites (blue) on the inner border and with an outside border containing 34 heptagonal sites (green) and 21 hexagonal sites (red); b) with 422 hexagonal sites (red), completed with 34 pentagonal sites (blue) on the inner border and with an outer border containing 55 heptagonal sites (green) and 34 hexagonal sites (red). Inner border have blue points inside, with steps of zig-zag between bluepoints containing one or two red points. This is labeled $l$ or $s$ in a). Outer border have red points on the outside with one or two green points between two red points. The inner and outer sequence of $l$ and $s$  are the same, starting at the first ``blue point'' or the first ``green point''. Three families of parastichies are clearly visible, they are characterized by three Fibonacci numbers $(f_{u-1},f_{u},f_{u+1})$ indicating the increase of indices jumping from a points to a neighboring point following a parastichies. These number are indicated on the figure (they correspond to $u=9$ for a) and $u=10$ for b) ). Notice that if we orient border following parastichies defined by the smaller increase given by $f_{u-1}$, the border in a) and b) are in reversed orientation: this is associated with the parity of $u$. }
\label{f11}
\end{figure}
%
%

The inner border of a ring is a zigzag line containing   pentagonal sites whose number is given by a Fibonacci number $f_{u-1}$ and hexagonal sites whose number is $f_{u}$. The figure~\ref{f11} gives two examples of hexagonal rings: one containing $134$ hexagons, characterized by the medium neighbor separation $\delta s=f_9=34$ the other with $422$ hexagons  and $\delta s=f_{10}=55$. Consider  such a ring  as a hexagonal lattice, with points on vertices of regular triangles, wrapped around a cylinder and with the primitive cell define by two vectors $\overrightarrow{a},\overrightarrow{b}$ defining elementary translations from a point to its  first neighbors  defined by $f_{u-1}$ and $f_{u}$. Then the ring can be seen as a multiple cell defined by two vectors expressed in the primitive cell by coordinates $(f_{u-1},-f_{u})$ and $(x,y)$ where $x$ and $y$ depend on the ring width. The area of the multiple cell, using the area of the  primitive cell as unit is given by the determinant:
$
\left[
\begin{array}{cc}
 f_{u-1} & x \\
 -f_{u} & y
\end{array}
\right]
 $. It is the number of points in the multiple cell.

 Wrapping of the multiple cell is obtained by identification of the translation $(f_{u-1},-f_{u})$ to the identity which result of gluing of two sides of this cell. The two other sides give the two boundaries of the ring. This type of description shows that the two boundaries are equivalent with the translation $(x,y)$. Counting the number of points in the ring comes to count the number of points in the cell but in the cell the two boundaries are equivalent, so both count for one. Then we have to add to the cell area the number of points on a border which is $f_{u-1}+f_{u}$. Following the number of points in the ring including pentagonal sites inside and heptagonal and hexagonal sites outside, as on the figure~\ref{f11} is given by the determinant
 $
\left[
\begin{array}{cc}
 f_{u-1} & x+1 \\
 -f_{u} & y+1
\end{array}
\right]
 $. In the example of figure~\ref{f11}--a this is
  $
\left[
\begin{array}{cc}
 21 & 0 \\
 -34 & 10
\end{array}
\right]
 $ with $x=-1$ and $y=9$ giving 210 sites in the ring. For figure~\ref{f11}--b
 it is
 $\left[
\begin{array}{cc}
 34 & 13 \\
 -55 & -5
\end{array}
\right]
$  with $x=12$ and $y=-6$ giving 545 site in the ring.
 %
%
%
\begin{figure}[tbp]
\includegraphics{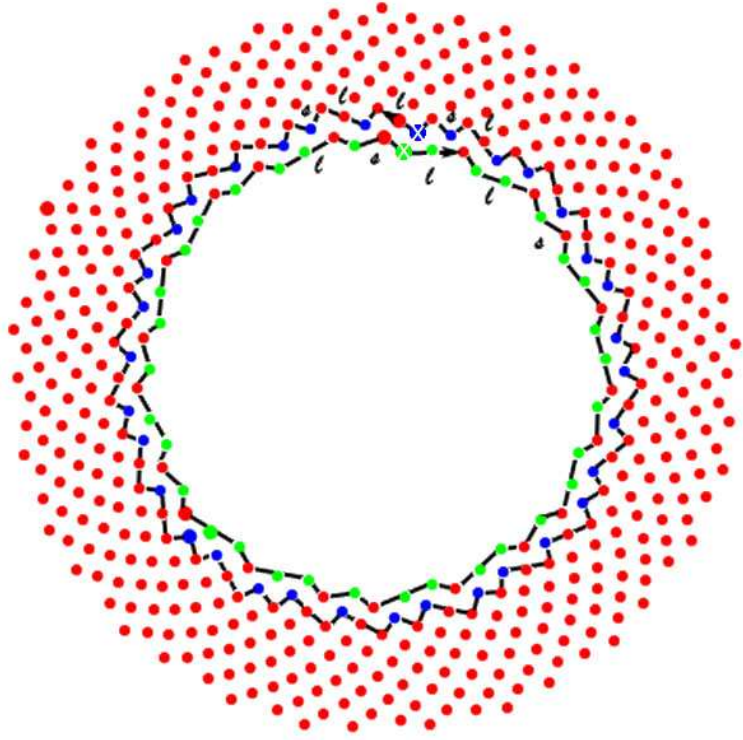}

\caption{The outside border of the ring with 210 sites and the inside border of the ring  with 545 sites. These boundaries are two chains of sites, one with heptagonal sites (green) separated by one or two hexagonal sites (red)have 55 sites, the other with pentagonal sites (blue) separated by hexagonal sites (red) have 89 sites. Starting from the first green site (cross) and rotating as indicated by arrows we can describe chains as a succession of long $(l)$ or short $(s)$ segments.}
\label{f12}
\end{figure}
%
%
\subsection{from boundary to boundary}

It is convenient to refer rings by the number $u$ in $f_u$ corresponding to their most visible parastichy. Their boundaries have $f_u+f_{u-1}=f_{u+1}$ points. How is related the outside border for $u$ with the inside border for $u+1$.
The figure~\ref{f12} show this relation with the example $u=9$ so with $f_u=34$ and $f_{u+1}=55$. A chain can be described as a chain of long segments (two heptagonal or pentagonal sites separated by two hexagonal sites) or short segment (two heptagonal or pentagonal sites separated by one hexagonal site). On this example the first chain is coded
 $LLSLLSLSLLSLLSLSLLSLS$ and the other is coded $LLSLLSLSLLSLLSLSLLSLSLLSLLSLSLLSLS$, but it is very important to notice that the chains are oriented in order to have an increase of site index, when jumping from their first site (first heptagon or pentagon) to the close one corresponding to the positive Fibonacci number $f_{u-1}$ for the first chain or $f_{u}$ for the other. Then the orientation of the two chains are reversed. Nevertheless as chains are palindromes using an other origin with a circular permutation, leads to symmetric representation.

The  inflation law relating the two chains is $L\rightarrow LS$ and $S\rightarrow L$,  as usual  for Fibonacci chains but this breaks the symmetry. It is then possible to have a law preserving the palindromic symmetry, but relating a chain defined by $u-1$ to a chain defined by $u+2$. This law which is $L\rightarrow LSLSL$ and $S\rightarrow LSL$ relate chains of lengthes $f_u$ and $f_{u+3}$. This law works perfectly well if lengthes are even numbers. If lengthes are odd numbers the two first letters have to be put at the sequence end to respect symmetry.

It is possible to conclude that the boundaries are related by symmetry operations. Inside a given ring the inner and the outer boundaries are related by some kind of translation, but which appears in the plane as a spiraling transformation coupling a rotation and a radial expansion. Then going to the next ring there is an inflation symmetry associated to a reverse of orientation.

\section{Noble phyllotaxis vs. false phyllotaxis}
\subsection{Other values of the parameter $\lambda$ and their approximants}

All the properties we have discussed above are a consequence of the continued fraction expansion of $\lambda$ which gives, by truncation, convergent rational approximants \cite{rivier}. When $\lambda$ is a noble number, $\{q_1,q_2, \ldots, 1,1,1,\ldots\}$ in continued fraction expansion it is possible to extrapolate from  $\lambda=1/\tau=\{1,1,1,1,\ldots\}$. False phyllotaxis is obtained with other rational or irrational $\lambda$, but never appear in nature. It is interesting to follow distances between neighbors (similarly to figure~\ref{f9} given for true phyllotaxis) as a function of $s$ and how they evolve from grain to  grain. Consider the three distances between a site at $s$ and  neighboring site at $s+\delta s_i$ as three functions of $s$ with constant $\delta s_i$ characteristic of each grain. These three functions have minimum  values which are  different, so leading to a dispersion of neighbor distances. Going from grain to grain so with   new sets of $\delta s_i$ two minimum are increasing but there is a   smallest distance with a minimum decreasing  with $s$. Voronoi cells, even if they remain hexagons in grains are more and more elongated with increasing $s$ leading to a spider web aspect.  This is the signature of the non-uniformity associated to non noble numbers.

In the case of a true phyllotaxis we have seen in paragraph~$3.2$ that neighbors are gathered on $f_u$ Fermat spirals (parastichies) when they are separated by $\delta s=f_u$. These is because convergent approximants of $\lambda=1/\tau$ obtained by truncation of it continued fraction expansion have the form $f_{u-1}/f_u$. In the case of non noble numbers, the truncation of the continued fraction expansion of $\lambda$ at $q_u$ leads  to principal rational approximant $\{\lambda^\prime= r_u/p_u; r_u,p_u\in  \mathbb{Z}\}$.
Then some neighbors are separated by $\delta s= p_u$ and gathered on $p_u$ similar spirals. Successive $p_u$ and $r_u$ are related by:
\begin{eqnarray}
p_u=q_u p_{u-1}+p_{u-2}
\label{equa5}.
\end{eqnarray}
In a grain consider a Delaunay triangle of close neighbors. The three edges of the triangle defined three directions of parastichies in the grain. The homogeneity of the grain require that these three families of parastichies, characterized by three index separations $\{\delta s_1,\delta s_2,\delta s_3\}$ are all the same in the grain. If the three separations are given in increasing order, the closure of edges of the triangle results in the triangular relation $\delta s_3=\delta s_1+\delta s_2$. If in a grain, a site $s$ is neighboring a site $s+p_u$ the triangular relation and equation \ref{equa5} are incompatible except if $q_u=1$ (as it is for all not to small $u$ in true phyllotaxis). To solve this incompatibility new intermediate approximants appear: to $r_u/p_u$ are associated $q_u$ approximants $r_u^{(i)}/p_u^{(i)}$ with the sub-index $(i)\in [1,q_u]$.  The integer $p_u^{(i)}$ and $r_u^{(i)}$ are defined by
$p_{u+1}^{(1)}=p_{u-1}+p_{u}$ and $p_{u+1}^{(i)}=p_{u+1}^{(i-1)}+p_{u}$, so we can write  $p_{u+1}^{(0)}=p_{u-1}$ and
$p_{u+1}^{(q_u)}=p_{u+1}$. Using  denominators of these intermediate approximants it is then possible to satisfy the triangular relation with the separations. In this case at least one principal approximant denominator $p_u$ appear in this list, and it appears in $q_u+2$ successive  grains.

Notice that distances between sites separated by a $\delta s$ related to an approximant $r_u^{(i)}/p_u^{(i)}$ are given by a similar relation to (\ref{equa3}) replacing $f_{u-1}$ and $f_u$ by $r_u^{(i)}$ and $p_u^{(i)}$. This can be done also in order to have minimum values using ( \ref{equa3b}).

\subsection{Phyllotaxis with noble number}
Phyllotaxis is characterized by two parameters. One, $\lambda$, describes the azimuthal (circle) map $\theta (s+1) = \theta (s) + 2\pi \lambda$, independently of the radial (growth) map $r(s)$. The radial map defines the generative spiral $r (\theta) $. On a flat substrate, the parameter $\alpha$ describes all kinds of parabolic spirals $r (\theta) = a \theta^{\alpha} $, but also the equiangular spiral $r (\theta) = a \exp {\theta}$ ($ \alpha = \infty$ ) and the dense spiral
 $r (\theta) = a \ln \{1 + \theta\}$ ($ \alpha = 0$) \cite{riviergoldar}. On a curved substrate, the spiral is defined by projection and an effective growth parameter $\alpha > 0$ can be defined, with $\alpha_0 = 1/2 $ for a flat substrate, $\alpha > \alpha_0 $ and a looser spiral for a substrate with positive curvature, and $\alpha < \alpha_0 $ and a tighter spiral for a substrate with negative curvature (our next paper).
In this paper, we concentrate on the ``ideal'' flat case, $\lambda = 1/\tau$, $\alpha_0 = 1/2 $ that yields a sequence of concentric circular grain boundaries.

An infinite phyllotactic structure can be obtained from any $0 < \lambda < 1$ that is a noble irrational, $\lambda = \{q_1,q_2, \ldots, q_u, 1/\tau\}$, beginning with  an arbitrary rational $ \{q_1,q_2, \ldots, q_u\}$ and terminating with a golden tail $1/\tau = \{1,1,\ldots,1,\ldots\}$ (continuous fraction expansion with $q_i \geq 1$). Now, all rationals between $0$ and $1$ can be arranged and ordered on a Farey tree, and the golden tail oscillates down the tree starting from its defining rational $ \{q_1,q_2, \ldots, q_u\}$, leading to a cascade of regular (parastichy) transitions through grain boundaries \cite{rothen,rivierkoch,levitov,rivier2,douadyJeanB,kochJeanB}. Note that the topological mechanism is always the same, articulated through the golden mean $1/\tau = \{1,1,\ldots,1,\ldots\}$. The actual value of the parameter $\lambda$ also oscillates between narrower bounds as the structure becomes larger. This implies also that the measurement of $\lambda$ from a given, finite structure is not that precise, and that it is collimated as the structure grows in size. Note that there are no exception in nature (ie no false phyllotaxis). Such tree already appears in Koch's thesis and in Rothen and Koch II\cite{rothen}, but the remark that it was isomorphic to Farey is due to Rivier\cite{rivierkoch}. It was made independently by Levitov \cite{levitov}.
Regular and singular transitions were first introduced by Koch and Rothen \cite{kochJeanB}. In Levitov, the singular transition branches are even disconnected (as an energy flow) from the main Farey tree\cite{leelevitov}.

The corresponding structure has a more or less disorganized central core moulded by the starting rational $ \{q_1,q_2, \ldots, q_u\}$, followed by the well established phyllotactic structure moulded by the golden tail. Obviously, the structure must be (much) larger than its core to exhibit it. Note that even for $\lambda = 1/\tau$ there is an apparent core for $s \leq 30$ within which  successive grain boundaries touch or even overlap. Topologically, the structure is perfect down to $s = 1$. (Table 1). Overlap occurs on strongly negative substrate (our next paper).

False phyllotaxis exhibit, beyond their core, spider-web structures, generated by $\lambda$ that are either rationals, $\lambda = \{q_1,q_2, \ldots, q_u,\infty\}$   or Liouville irrationals $\lambda = \{q_1,q_2, \ldots, q_u, Q, \ldots \}$, with the integer $Q  \gg q_i \geq 1$. On the Farey tree, the Liouville irrational goes through a large number $ \leq Q$ of singular transitions.

\section{Conclusions}

An infinite number of sites at distance $d$ organized on the nodes of a triangular tiling defined by a hexagonal lattice have the translational and rotational symmetries of this lattice. To each nodes it is possible to associate an hexagon considered as  its Voronoi cell. This tessellation with regular hexagonal cells, obeys the Euler relation
$F-E + V = 0$ where $F, E$ and $V$ are respectively the number of faces, edges and vertices in the tiling. If we consider a limited number of points occupying a finite domain limited by a circular boundary, the Euler relation  becomes $F-E + V = 1$ and non-hexagonal cells appear among the hexagonal cells. The distribution of points in the disk therefore contains inherent defects and is no longer invariant under the symmetry operations of classical crystallography.

The algorithm of phyllotaxis  builds a dense organization of points in a  disk that optimizes the homogeneity of the area associated to each point and the isotropy of the environment of each point in a situation of a circular symmetry. To identify the laws specific to this type of organization we have described it using Voronoi  and Delaunay decompositions which make apparent the presence of intrinsic defects.
These defects are pentagonal and heptagonal cells associated two by two in dipoles organized along narrow concentric rings separating larger rings containing hexagonal cells continuously deformed. Such an organization can be understood by considering two types of disorder in the organization: a  metric disorder corresponding to fluctuations of distances between  first neighbors, as the hexagonal cells are not regular, and a topological disorder corresponding to the presence of  pentagonal and heptagonal cells associated into dipoles. These two disorders interact to build the organization.
Notice that as usual,  dislocations are called topological defects, but the notion of defect refer to the perfect crystalline structure. In the phyllotaxis pentagon-heptagon dipoles are a basic constituent of the structure.

The role of  pentagonal and heptagonal cells dipoles is to introduce rows of new cells needed to maintain the homogeneity of area when the radius of the structure increases, as does a dislocation in metallurgy. This organization is reminiscent also of grains and grain boundaries of dislocations in metals, but here the grain boundaries are imposed by the circular symmetry whereas in  metal grain boundaries they result from processing treatments. Finally a very specific order is observed   in the  organization of these dipoles in concentric grain boundaries:   the
radius of the grain boundaries,  or the width  of grains,  follow  the Fibonacci series, while
the  sequence of   dipoles along the grain boundaries are those of one-dimensional quasicrystalline sequences which are deduced from each other by an inflation-deflation rule which leads to a self-similar structure.

Thus the topological constraint of the circular symmetry introduces an original inflation-deflation symmetry replacing the translational and rotational symmetries of classical crystallography. Still using the language of metallurgy, the Voronoi cell and parastichies could be seen as the repeat unit and the crystallographic lattice planes of this unconventional  structure.

Rings of dipoles, the grain boundaries, with their self-similar organization are the fundamental characteristic of phyllotaxis. Presently we are investigating phyllotactic tilings on the sphere and on the hyperbolic plane, always keeping a constant area for each site. The geometry of the dipole rings and their succession is the same as on the flat disk. All the effect of the Gaussian curvature of the tiled surface only concerns the width of the hexagonal grains. So whatever is the underlying geometry grain boundaries are structurally blocked, without any adjustable parameter. This is true with a topological but also a metric point of view.

The organization of the florets of the  flowers of many plants is the earliest known example of implementation of the algorithm of phyllotaxis, it was indeed at the origin of  the development of this domain of studies. As already quoted in the introduction recent works on purely physical systems  offer new examples   \cite{rivier,douady,yoshikawa}.

All these examples are macroscopic assemblies of elements such as florets, leaves, convection cells, ferrofluid droplets and bubbles. They have in common, not only a circular symmetry, but also the ability of their elements to move with respect to each other in order to adjust their distribution in the curse of the growth process. Those symmetry and behavior are also presented at the microscopic level by molecular associations of soft condensed matter and biological materials. Their molecular interactions allow an internal mobility and their interfacial tension in solution imposes the circular symmetry so that they could be relevant of a similar approach. This is why we proposed it for describing collagen fibers \cite{charvolinsadoc}. If this were so, whereas in the cases of flowers or droplets the growth is generated from the center and in that of the convection cells the organization builds itself at a collective transition, this growth of fibrils by addition at their surface would provides a third example of the power of the algorithm of phyllotaxis in structural studies.

\end{document}